\documentclass[letterpaper,titlepage,11pt]{article}

\pdfoutput=1

\usepackage[T1]{fontenc}
\usepackage{amssymb,amsmath,amsfonts}
\usepackage{mathrsfs,mathtools}
\usepackage{ccaption}
\usepackage{graphicx}
\usepackage{tikz}
\usetikzlibrary{decorations.pathmorphing,calc}

\usepackage[
      colorlinks=true,
      linkcolor=blue,
      urlcolor=magenta,
      filecolor=green,
      citecolor=red,
      pdfstartview=FitV,
      pdftitle={},
        pdfauthor={Arjun Bagchi, Daniel Grumiller, Jakob Salzer, Sourav Sarkar, Friedrich Scholler},
        pdfsubject={},
        pdfkeywords={},
        pdfpagemode=UseNone,
        bookmarksopen=true
      ]{hyperref}

\usepackage[all]{hypcap} 

\def\clock{{\count0=\time
           \divide\count0 60
           \ifnum\count0<10 0\fi\the\count0
           \multiply\count0 -60 \advance\count0 \time
           :\ifnum\count0<10 0\fi \the\count0
         }}
\newcommand{\timestamp}{{\small\vbox{\hbox{\tt\jobname.tex}
\hbox{\the\day/\the\month/\the\year, \clock}}}}

\usepackage{amsmath,amssymb}
\usepackage{verbatim}
\usepackage{graphicx}
\usepackage{hyperref}
\usepackage{color}

\DeclareFontFamily{OT1}{rsfs}{}
\DeclareFontShape{OT1}{rsfs}{m}{n}{ <-7> rsfs5 <7-10> rsfs7 <10->rsfs10}{} 
\DeclareMathAlphabet{\mycal}{OT1}{rsfs}{m}{n}

\newcommand{\be}[1]{ \begin{equation}\label{#1} }
\newcommand{\ee}{\end{equation}}
\newcommand{\bea}[1]{\begin{eqnarray}\label{#1} }
\newcommand{\eea}{\end{eqnarray}}

\renewcommand{\phi}{\varphi}

\newcommand{\eq}[2]{\begin{equation} #1 \label{#2} \end{equation}}

\DeclareMathOperator{\extdm}{d}
\newcommand{\extd}{\extdm \!}

\begin{document}



\begin{titlepage}
\leftline{}{TUW--14--10}
\vskip 3cm
\centerline{\LARGE \bf Flat space cosmologies in two dimensions}
\vskip 2mm
\centerline{\Large \bf Phase transitions and asymptotic mass-domination} 
\vskip 1.6cm
\centerline{\bf Arjun Bagchi$^{a,b}$, Daniel Grumiller$^{b}$, Jakob Salzer$^{b}$,} 
\centerline{\bf Sourav Sarkar$^{a}$, and Friedrich Sch\"oller$^{b}$}
\vskip 0.5cm
\centerline{\sl $^{a}$Indian Institute of Science Education and Research, Pune 411008, India}
\smallskip
\centerline{\sl $^{b}$Institute for Theoretical Physics, Vienna University of Technology}
\centerline{\sl Wiedner Hauptstrasse 8-10/136, A-1040 Vienna, Austria}

\vskip 0.5cm
\centerline{\small E-mail: \tt \href{mailto:a.bagchi@iiserpune.ac.in}{a.bagchi@iiserpune.ac.in}; \href{mailto:grumil@hep.itp.tuwien.ac.at}{grumil}, \href{mailto:salzer@hep.itp.tuwien.ac.at}{salzer}, \href{mailto:schoeller@hep.itp.tuwien.ac.at}{schoeller@hep.itp.tuwien.ac.at};}
\centerline{\small \tt  \href{mailto:ssarkar@students.iiserpune.ac.in}{ssarkar@students.iiserpune.ac.in}}

\vskip 1.6cm
\centerline{\bf Abstract} \vskip 0.2cm \noindent

We study flat space cosmologies in two dimensions by taking the flat space limit of the Ach\'ucarro--Ortiz model. We unravel a phase transition between hot flat space and flat space cosmologies, and derive a new dilaton-dependent counterterm required for the consistency of the Euclidean partition function. Our results generalize to asymptotically mass-dominated 2-dimensional dilaton gravity models, whose thermodynamical properties we discuss. The novel case of asymptotic mass-domination is neither covered by the comprehensive discussion of \href{http://www.arXiv.org/abs/hep-th/0703230}{{\tt hep-th/0703230}} nor by the more recent generalization to dilaton gravity with confining $U(1)$ charges in \href{http://www.arXiv.org/abs/1406.7007}{{\tt 1406.7007}}.

\end{titlepage}
\pagestyle{empty}
\small
\normalsize
\newpage
\pagestyle{plain}
\setcounter{page}{1}


\addtocontents{toc}{\protect\setcounter{tocdepth}{3}}

\tableofcontents

\renewcommand*\listtablename{Table and Figure}
\renewcommand*\listfigurename{}

\listoftables

\vspace*{-1.5truecm}

\listoffigures

\newpage

\section{Introduction}\label{se:1}

Einstein gravity in three dimensions is special due to the absence of propagating physical degrees of freedom \cite{Deser:1983tn,Witten:1988hc}. It was thus a great surprise when Ba{\~{n}}ados, Teitelboim and Zanelli (BTZ) discovered that 3-dimensional Anti-de~Sitter (AdS$_3$) spacetimes have black hole solutions \cite{Banados:1992wn}. These BTZ black holes are locally AdS$_3$ and are perhaps best described as orbifolds of AdS$_3$ \cite{Banados:1992gq}. 
In 3-dimensional (3d) flat space there are no black holes, but there exist analogues of the BTZ black holes, which turn out to be cosmological solutions and are called Flat Space Cosmologies (FSCs). These can be obtained in the flat space limit of non-extremal rotating BTZ black holes, where the outer event horizon of the black hole is pushed out to infinity. The cosmological solution turns out to be the remnant of the region between the two BTZ Killing horizons. FSCs can also be viewed as appropriate orbifolds of 3d flat space \cite{Cornalba:2002fi,Cornalba:2003kd}.  

One of the very interesting features of the BTZ black holes is that there exist Hawking--Page phase transitions \cite{Hawking:1982dh}, which take thermal AdS$_3$ to BTZ black holes. These transitions also exist in higher dimensions and are not purely 3d phenomena.  

Recently a new type of phase transition was discovered: 3d flat space, if heated up sufficiently and stirred gently, undergoes a phase transition to a FSC \cite{Bagchi:2013lma}. In spirit, this phase transition is similar to the above mentioned Hawking--Page phase transition. But there is an important difference: this is probably the first example of a phase transition between a time-independent and a time-dependent solution. The phase transition thus provides a unique way of creating a time evolving universe by heating ordinary flat space, albeit so far only in three dimensions.

It is interesting to find out whether that phase transition is a specific property of three dimensions or, like the Hawking-Page transition, it persists in other dimensions. In this paper we take a first modest step towards showing the generality of the phase transition by proving that a similar kind of phase transition exists in two dimensions. 

To set the stage, we summarize now briefly some relevant results in three dimensions. Flat space Einstein gravity, $R_{\mu\nu}=0$, has a two parameter family of FSCs \cite{Cornalba:2002fi,Cornalba:2003kd}, whose line-element in Euclidean signature
\begin{equation}
  \extd s^2 = r_+^2\,\big(1-\tfrac{r_0^2}{r^2}\big)\,\extd \tau^2+\frac{\extd r^2}{r_+^2\,\big(1-\frac{r_0^2}{r^2}\big)}+r^2\,\Big(\extd \varphi-\frac{r_+r_0}{r^2}\extd \tau\Big)^2  \label{FSC3}
\end{equation}
is subject to the identifications
\eq{
(\tau,\,\varphi)\sim(\tau+\beta,\,\varphi+\beta\,\Omega)\sim(\tau,\,\varphi+2\pi)
}{eq:intro1}
with inverse temperature $\beta = 2\pi r_0/r_+^2$ and angular potential $\Omega=r_+/r_0$.
The inverse Wick rotation $\frak t=i\tau$, $\hat r_+=-ir_+$, together with the coordinate transformation $\hat r_+ \frak t =x$, $r_0\varphi=x+y$, $(r/r_0)^2=1+(E{\mathfrak t})^2$, where $E=\hat r_+/r_0$, yields the Minkowskian version of FSCs.
\eq{
\extd s^2=-\extd{\mathfrak t}^2+\frac{(E{\mathfrak t})^2}{1+(E{\mathfrak t})^2}\,\extd x^2+ \big(1+(E{\mathfrak t})^2\big)\,\Big(\extd y+\frac{(E{\mathfrak t})^2}{1+(E{\mathfrak t})^2}\,\extd x\Big)^2
}{eq:intro3}
The line-elements \eqref{eq:intro3} comprise a two parameter family of locally flat time-dependent exact solutions of the vacuum Einstein equations \cite{Barnich:2012xq,Bagchi:2012xr}. For positive (negative) $\mathfrak t$ they describe expanding (contracting) universes from (towards) a cosmological horizon at $\mathfrak t=0$. The parameter $E$ corresponds physically to the temperature associated with the FSC \cite{Cornalba:2002nv}. Note that the Euclidean time $\tau$ (Minkowskian time $\mathfrak t$) essentially corresponds to the spatial coordinate $x$ (radial coordinate $r$).

Our main idea is to Kaluza--Klein reduce the Euclidean metric \eqref{FSC3} and the associated Einstein--Hilbert action along the $\varphi$-direction. For non-vanishing $r_+$ and $r_0$ there is a $\varphi$-twist in the $\tau$-identification [see the first identification \eqref{eq:intro1}], which amounts to the presence of a $U(1)$-gauge field in the Kaluza--Klein split. Moreover, the overall $r^2$-factor in the last term of the line-element \eqref{FSC3} means that there is a non-constant scalar field in the Kaluza--Klein split, namely the dilaton.  Thus, we should expect the Kaluza--Klein reduction of the 3d Einstein--Hilbert action to yield a specific 2-dimensional (2d) Einstein--Maxwell--Dilaton theory. Indeed, we are going to show that this is precisely what happens, but as we shall demonstrate there is a number of subtleties, particularly in the presence of (asymptotic) boundaries. 

After establishing the existence of 2d FSCs by constructing the flat-space limit of the Ach\'ucarro--Ortiz model \cite{Achucarro:1993fd} we shall reach one of our main aims, namely to prove the existence of a phase transition between hot flat space (HFS) and 2d FSCs.

We then go on and generalize our results to a large class of 2d Einstein--Maxwell--Dilaton models, namely models whose solutions have metrics that asymptotically are dominated by the mass term. We shall introduce and carefully discuss the notion of asymptotic mass-domination. Finally, we intend to verify which features of the flat space Ach\'ucarro--Ortiz model remain intact and under which conditions, for instance the existence of a phase transition or the positivity of specific heat and electric susceptibility.

This paper is organized as follows. 
In section \ref{se:2} we derive the flat-space limit of the Ach\'ucarro--Ortiz model, including its boundary terms, and study its thermodynamics, including phase transitions between HFS and FSC.
In section \ref{se:3} we introduce the notion of asymptotic mass-domination and generalize the results of the previous section to asymptotically mass-dominated 2d dilaton gravity with non-confining $U(1)$ charge. In particular, we prove generically the existence of a phase transition between HFS and (generalized) FSC, subject to an inequality.

Before starting we mention some of our conventions, which are compatible with the conventions of \cite{Grumiller:2007ju}. We work in Euclidean signature unless mentioned otherwise; when using Minkowskian notions such as `Killing horizon' we always imply that the corresponding analytically continued entity has this property. Our sign conventions for the Ricci tensor are fixed by $R_{\mu\nu}=\partial_\alpha\Gamma^\alpha{}_{\mu\nu}+\dots$. To reduce clutter we fix the 2d Newton constant as $8\pi G_2 = 1$. For manifolds $\mathcal M$ with boundary $\partial \mathcal M$ we denote the trace of extrinsic curvature by $K$ and the determinant of the induced metric on the boundary by $\gamma$.

\section{Phase transition of flat space in two dimensions}\label{se:2}

In this section we consider a specific model where the existence of a phase transition is plausible on general grounds. The main purpose is to show that this phase transition indeed exists, and to collect a few technical and conceptual insights on the way that we intend to generalize in the next section.

In section \ref{se:2.0} we recall aspects of FSC in three dimensions and their phase transition to flat space.
In section \ref{se:2.1} we perform a Kaluza--Klein reduction to two dimensions of bulk and boundary terms, thereby providing a well-defined variational principle for the flat space Ach\'ucarro--Ortiz model.
In section \ref{se:2.2} we calculate the Euclidean partition function and unravel the phase transition.
In section \ref{se:2.3} we study thermodynamical properties of the flat space Ach\'ucarro--Ortiz model.

\subsection{Flat space (cosmologies) and their phase transition}\label{se:2.0}

We collect now further salient features of 3d FSCs \eqref{FSC3}. A reader familiar with FSCs can skip to section \ref{se:2.1}.

As mentioned in the introduction, FSC spacetimes are the flat space analogues of non-extremal rotating BTZ black holes \cite{Banados:1992wn}. Given the interpretation of the latter as orbifolds of Anti-de~Sitter (AdS) \cite{Banados:1992gq}, it is not surprising that FSCs have a similar interpretation, namely as shifted-boost orbifolds of $\mathbb{R}^{1,2}$ \cite{Cornalba:2002fi,Cornalba:2003kd}.

In fact, many properties of FSCs can be deduced as a cosmological constant to zero limit from corresponding BTZ results, though the limit is often subtle and not always accessible, at least not straightforwardly. For example, on the bulk side \cite{Barnich:2010eb,Bagchi:2010zz,Bagchi:2012cy,Barnich:2012aw} the symmetries of asymptotically AdS spacetimes are contracted to the Bondi--van~der~Burg--Metzner--Sachs (BMS) algebra \cite{Bondi:1962,Sachs:1962,Ashtekar:1996cd}. On the field theory side, the conformal field theory symmetries are contracted to Galilean conformal symmetries, previously considered in the context of non-relativistic limits of conformal field theories \cite{Bagchi:2009my}, though in the present context one should rather think of the contraction as an ultra-relativistic limit. 

Also other gauge/gravity aspects have flat space analogues. For instance, the microstate counting of BTZ black holes \cite{Strominger:1997eq} has a corresponding flat space analogue \cite{Barnich:2012xq,Bagchi:2012xr} that matches the Bekenstein--Hawking entropy of FSCs with the asymptotic growth of states in the putative dual Galilean conformal field theory.

We state now some thermodynamical properties of FSCs, starting with the definitions of mass $M$ and angular momentum $J$.
\begin{equation}
  M = \frac{r_+^2}{8G} \qquad J=-\frac{r_+r_0}{4G} 
\end{equation}
These quantities emerge as zero-mode boundary charges from a canonical analysis of flat space Einstein gravity with flat space boundary conditions \cite{Barnich:2006av,Bagchi:2012yk}. The free energy $F=-M$ satisfies a first law
\eq{
\extd F(T,\,\Omega) = -S\,\extd T-J\,\extd\Omega
}{eq:phase39}
with the Bekenstein--Hawking entropy
\eq{
S = \frac{2\pi r_0}{4G}\,.
}{eq:phase40}
The specific heat at constant angular potential is always positive, $C=S=\pi^2T/(G\Omega)^2$, and, just like entropy, satisfies a version of the third law (specific heat and entropy both vanish at zero temperature). 

Expressed as function of its natural variables, the FSC free energy reads
\eq{
F(T,\,\Omega) = T\,\Gamma = -\frac{\pi^2 T^2}{2G\Omega^2}\,,
}{eq:phase41}
where $\Gamma$ is the on-shell Einstein--Hilbert action evaluated on the corresponding FSC solution.
By contrast, the free energy of HFS, $\extd s^2=\extd\tau^2+\extd r^2+r^2\,\extd\varphi^2$ with the identifications \eqref{eq:intro1}, is always constant, $F(T,\,\Omega)=-1/(8G)$. Therefore, there is a phase transition at the critical temperature \cite{Bagchi:2013lma}
\eq{
T_c = \frac{\Omega}{2\pi}\,.
}{eq:phase42}
So in the high temperature regime, $T>T_c$, the preferred Euclidean saddle point is FSC, while in the low temperature regime, $T<T_c$, the preferred Euclidean saddle point is HFS.


\subsection{Kaluza--Klein reduction \'a la Ach\'ucarro--Ortiz}\label{se:2.1}

In order to derive the free energy from the on-shell action we have to make sure that the action has a well-defined variational principle for a given set of boundary conditions, in the sense that the first variation of the action vanishes when evaluated on any classical solution compatible with the boundary conditions.
In \cite{Detournay:2014fva} it was shown that the action
\begin{equation}
  \Gamma =
    - \frac{1}{16\pi G_3} \int_{\mathrlap{\mathcal M}} \; \extd^3 x \sqrt{g}\, R
    - \frac{\alpha}{8\pi G_3} \int_{\mathrlap{\partial \mathcal M}} \; \extd^2 x \sqrt{\gamma}\, K \label{flatgamma3}
\end{equation}
with $\alpha=1/2$ yields a well-defined variational principle for flat space boundary conditions in three dimensions. 
Note that the usual Gibbons--Hawking--York (GHY) boundary term would require $\alpha=1$, which is inconsistent with flat space boundary conditions. The action \eqref{flatgamma3} with $\alpha=1/2$ was used to determine the FSC free energy \eqref{eq:phase41}.

Since the action \eqref{flatgamma3} yields a well-defined variational principle one expects that by dimensional reduction to two dimensions a well-defined 2d action principle can be obtained. This is shown explicitly in the following by performing a Kaluza--Klein reduction analogue to the one performed by Ach\'ucarro and Ortiz \cite{Achucarro:1993fd}. 

We use the following conventions: 3d quantities are discriminated from 2d quantities by a tilde, i.e., $\tilde{R}$ denotes the Ricci scalar of the 3d manifold, while $R$ is the Ricci scalar of the 2d manifold. 3d (2d) spacetime indices are $A, B, C, \dots$ ($a, b, c, \dots$). In the Cartan formalism 3d (2d) flat indices are denoted by $I, J, K, \dots$ ($i, j, k, \dots$). 

In terms of dreibeins the 3d and 2d metrics are written as
\begin{equation}
  \tilde{g}_{AB} = \tilde{e}_A^{I}\,\tilde{e}_B^{J}\,\delta_{IJ} \qquad
  g_{ab} = e_a^i\,e_b^j\,\delta_{ij} \, .
\end{equation}
In order to be able to perform the Kaluza--Klein reduction we assume cylindrical symmetry, i.e., the theory does not depend on the angular coordinate denoted by $\varphi$. Then, with no loss of generality we require that all but one of the dreibeins are orthogonal to the $\partial_{\varphi}$-direction. Thus, we set
\begin{equation}
  \tilde{e}^{i} = e^{i} \qquad i=1,2\,, \label{threeistwo}
\end{equation}
while the third dreibein, $\tilde{e}^{3}$, has a $\varphi$-component that we determine below. 

It is convenient to parametrize the 3d metric $\tilde{g}_{AB}$ (and analogously its inverse $\tilde{g}^{AB}$) in the following way
\begin{equation}
  \tilde g_{AB} =
    \begin{pmatrix}
      g_{ab} + X^{2} A_a A_b \phantom{\Big(} & \phantom{\Big(} X^{2} A_a \\
      X^{2} A_b \phantom{\Big(} &\phantom{\Big(} X^2
    \end{pmatrix}
\qquad 
 \tilde g^{AB} =
  \begin{pmatrix}
    g^{ab} \phantom{\Big(}&\phantom{\Big(} - A^a \\
    - A^b \phantom{\Big(}&\phantom{\Big(} X^{-2} + A_c A^c
\end{pmatrix}\,.
\label{eq:KK}
\end{equation}
With the assumption \eqref{threeistwo}, the remaining dreibein $\tilde{e}^3$ is therefore fixed to
\begin{equation}
\tilde{e}^3=X\left(A+\extd \varphi\right),
\end{equation}
where $A=A_a\extd x^{a}$ is the gauge field 1-form and the $x^{a}$ are the remaining coordinates, denoted by $\tau$, $r$. 

Since both connection one-forms $\tilde{\omega}^I{}_J$ and $\omega^i{}_j$ associated with the respective vielbeins are torsionless and metric compatible ($\tilde{\omega}^{IJ} = -\tilde{\omega}^{JI}$), we exploit Cartan's structure equations
\eq{
  \extd \tilde{e}^{I} + \tilde{\omega}^I{}_J\wedge \tilde{e}^{J} = 0
  \qquad \tilde{R}^I{}_J = \extd \tilde{\omega}^I{}_J+\tilde{\omega}^I{}_K\wedge \tilde{\omega}^K{}_J
}{eq:phase1}
to relate the connections and the respective curvatures.
Thus, we arrive at
\begin{subequations}
\label{eq:phase3}
\begin{align}
  \tilde R &= R - \frac{1}{4} X^2 F_{ab} F^{ab} - 2 X^{-1} \nabla^2 X\, \\
  \tilde K &= K + X^{-1} n^a \partial_a X,
\end{align}
\end{subequations}
where the unit normal vector $n^a$ points in the $\partial_r$-direction and $F_{ab}=\partial_aA_b-\partial_bA_a$. The metric and boundary metric determinants are related by
\eq{
  \sqrt{\tilde g} = X \,\sqrt{g} \qquad
  \sqrt{\tilde \gamma} = X \,\sqrt{\gamma}\,.
}{eq:lalapetz}

The relations \eqref{eq:phase3}, \eqref{eq:lalapetz} allow to Kaluza--Klein reduce the action \eqref{flatgamma3} to a 2d Einstein--Maxwell--Dilaton action:
\begin{equation}
\boxed{
\phantom{\Bigg(}
  \Gamma =
    - \frac{1}{2} \int_{\mathrlap{\mathcal M}} \; \extd^2 x \sqrt{g} \,\Big( X R - \frac{1}{4} X^3 F_{ab} F^{ab}\Big)
    - \frac{1}{2} \int_{\mathrlap{\mathcal \partial M}} \; \extd x \sqrt{\gamma}\, \Big( X K - n^a \partial_a X \Big) 
\phantom{\Bigg(}
}
\label{reducedaction}
\end{equation}
Here we set $2\pi G_2=G_3$ and used the convention $8\pi G_2=1$ mentioned at the end of our introduction. The action \eqref{reducedaction} is our first key result. It is a specific Einstein--Maxwell--Dilaton model that we call ``flat space Ach\'ucarro--Ortiz'' model.

It is now a simple exercise to show that the variational principle is well-defined for the action \eqref{reducedaction} with Euclidean flat space boundary conditions.
\begin{subequations}
 \label{eq:bc}
\begin{align}
 g_{ab}  &= \begin{pmatrix}
       g_{\tau\tau} = M + o(1) & g_{\tau r} = o(1) \\
       & g_{rr} = 1/M + o(1)
      \end{pmatrix} \label{eq:bc1} \\
 A_a &= \begin{pmatrix}
       A_\tau = o(1) \\
       A_r = o(1)
      \end{pmatrix} \label{eq:bc2} \\
 X & = r + o(1) \label{eq:bc3}
\end{align}
\end{subequations}
Note that the parameter $M$ in \eqref{eq:bc1} is allowed to fluctuate; the expression $o(1)$ means that the corresponding quantity tends to zero as the radial coordinate $r$ tends to infinity.
We shall be more explicit about the variational principle in section \ref{se:3} and merely state here the final result.
\eq{
\delta\Gamma\big|_{\textrm{\tiny EOM}} = 0
}{eq:phase2} 
The left hand side, $\delta\Gamma$, contains any variation of the fields that is compatible with the boundary conditions \eqref{eq:bc}. The subscript `EOM' denotes that after variation on-shell relations have been used, which we discuss in the next subsection.

\subsection{Phase transition in two dimensions}\label{se:2.2}

The thermodynamics of spacetimes is studied most conveniently in the saddle-point approximation to the Euclidean path integral \cite{Gibbons:1976ue}. 
The Euclidean partition function is given by
\begin{equation}
  Z \sim \sum_{\mathclap{g_{cl},\,A_{cl},\,X_{cl}}} \exp{\left(-\Gamma\left[g_{cl},\,A_{cl},\,X_{cl}\right]\right)} \times Z_{\textrm{\tiny Gauss}} \times Z_{\textrm{\tiny ho}}\label{pathfluc}
\end{equation}
with
\begin{equation}
  Z_{\textrm{\tiny Gauss}} = \int \mathcal{D}\delta g\, \mathcal{D}\delta\! A\, \mathcal{D}\delta X\, \exp{\left(-\tfrac{1}{2}\delta^2\Gamma\left[g_{cl},\,A_{cl},\,X_{cl};\delta g,\,\delta A,\,\delta X\right]\right)}\label{eq:angelinajolie}
\end{equation}
and higher order corrections contained in $Z_{\textrm{\tiny ho}}$. The sum in \eqref{pathfluc} extends over all classical solutions $g_{cl}$, $A_{cl}$, $X_{cl}$ compatible with the boundary conditions that we impose in order to evaluate the path integral. The choice of boundary conditions simultaneously specifies the thermodynamic ensemble and the appropriate thermodynamic potential $Y$ that depends on the quantities held fixed at the boundary \cite{Brown:1989fa}. This potential to leading order is then given by
\begin{equation}
  Y =  - T\,\ln Z \approx T \,\Gamma\big[\hat g_{cl},\,\hat{A}_{cl},\,\hat X_{cl}\big] \label{freebee}
\end{equation}
where $\hat g_{cl}$, $\hat A_{cl}$, $\hat X_{cl}$ denotes the most dominant Euclidean saddle point. In the present work we neglect corrections from Gaussian (or higher order) fluctuations as well as instanton corrections from sub-dominant saddle-points.\footnote{Leading corrections to the saddle-point approximation have been calculated recently for the entropy of 3d FSCs in \cite{Bagchi:2013qva}.}

As discussed e.g.~in \cite{Grumiller:2007ju,Detournay:2014fva} the saddle-point approximation to the Euclidean path integral is well-defined only if the action has a well-defined variational principle. Since we have achieved this in the previous subsection, we proceed now with a discussion of on-shell relations.

The equations of motion (EOM) descending from the action \eqref{reducedaction} are given by
\begin{align}
 \nabla_a \partial_b X - g_{ab} \nabla^2X &= -\tfrac12\,X^3\,\big(F_a{}^c F_{bc} - \tfrac14\,g_{ab} F^{cd} F_{cd}\big) \label{EOM1} \\
 R &= \tfrac34\, X^2 F^{ab} F_{ab} \label{EOM2} \\
 0 &= \nabla_{a}\big( X^3 F^{ab} \big)  \label{EOM3}
\end{align}
Integrating the last equation leads to a conserved $U(1)$ charge $q$,
\eq{
F_{ab} = \frac{2q}{X^3}\,\epsilon_{ab}\,.
}{eq:phase4}
Solving the other two EOM yields, in diagonal gauge, the line-element
\eq{
\extd s^2 = \big(M-\tfrac{q^2}{r^2}\big)\,\extd\tau^2 + \frac{\extd r^2}{M-\frac{q^2}{r^2}}
}{eq:phase5}
in axial gauge the Maxwell field
\eq{
A_\tau = \frac{q}{r^2}+{\rm const.} \qquad A_r = 0 
}{eq:phase6}
and the dilaton field
\eq{
X = r\,.
}{eq:phase7}
Note that all solutions \eqref{eq:phase5}-\eqref{eq:phase7} respect the flat-space boundary conditions \eqref{eq:bc}. We call such solutions ``2d flat space cosmologies'', since they yield exactly the line-element \eqref{FSC3} when inserted into the Kaluza--Klein ansatz \eqref{eq:KK}.\footnote{Note that we could explicitly identify angular momentum with charge $J=q$ and angular potential with electric potential $\Omega=\Phi$, but we choose to stick to the intrinsically 2d quantities $q,\Phi$. For a comparison of 2d and 3d quantities see table \ref{tab:1} below. } Moreover, as we show in appendix \ref{app:A}, the Penrose diagram also coincides with appropriate 2d sections of the Penrose diagram for 3d FSCs. The parameter $M>0$ in the line-element \eqref{eq:phase5} is a second constant of motion and physically corresponds to the mass of the solution.

Our first task is to determine the horizon and to fix the temperature, $T=\beta^{-1}$, which is done in the usual way by demanding that there be no conical singularity, so that the periodicity of Euclidean time, $\tau\sim\tau+\beta$, is fixed. We obtain
\eq{
T = \frac{q^2}{2\pi X_h^3}
}{eq:phase8}
where $X_h$ is the dilaton field evaluated at the horizon.
\eq{
X_h = r_h = \frac{|q|}{\sqrt{M}}
}{eq:phase9}

Our second task is to choose our thermodynamic ensemble. We elect to keep fixed the temperature $T$ and the electric potential $\Phi$. The rationale behind these choices is that they correspond precisely to the same choices in the related 3d discussion \cite{Bagchi:2013lma}. 
Physically, fixing $T$ and $\Phi$ at the boundary can be achieved by coupling the system asymptotically to some heat bath at constant temperature $T$ and constant electric potential $\Phi$ (with respect to some fiducial grounding). In the 3d interpretation the constant electric potential corresponds to a constant angular potential.

The electric potential between a cut-off surface $r=r_c$ and the horizon $r=r_h$ is given by
\eq{
\Phi(r_c) = A_\tau (r_c) - A_\tau(r_h) \,.
}{eq:phase11}
In the limit of interest the cut-off surface tends to the asymptotic boundary, $r_c\to\infty$, and the electric potential simplifies to
\eq{
\Phi = \lim_{r_c\to\infty}\Phi(r_c) = -\frac{q}{X_h^2}\,.
}{eq:phase12}

Finally, inserting the on-shell results above into the action \eqref{reducedaction} yields
\eq{
Y(T,\,\Phi) = -2\pi^2\,\frac{T^2}{\Phi^2}\,.
}{eq:phase10}
The result \eqref{eq:phase10} coincides precisely with the 3d result for free energy \cite{Bagchi:2013lma} (note that their $G_3=\tfrac14$ here).

We can now unravel the existence of a phase transition between HFS,
\eq{
\extd s^2_{\textrm{\tiny HFS}} = \extd\tau^2+\extd r^2\qquad A_\tau=A_r=0\qquad X=r
}{eq:phase17}
and 2d FSC \eqref{eq:phase5}-\eqref{eq:phase7}.
Evaluating free energy \eqref{eq:phase10} for the former yields
\eq{
Y_{\textrm{\tiny HFS}}(T,\,\Phi) = -\frac12\,.
}{eq:phase18}
While the right hand side does not depend on temperature or electric potential, we stress that any values of $T$ and $\Phi$ are consistent for HFS. Temperature is fixed again by setting the periodicity appropriately, $\tau\sim\tau+\beta$, with $\beta=T^{-1}$, and the electric potential can be set to any constant value.

Comparing the free energies \eqref{eq:phase10} and \eqref{eq:phase18} of the two admissible saddle points, we find that they are equal at the critical temperature
\eq{
T_c = \frac{\Phi}{2\pi}\,.
}{eq:phase19}
For a given electric potential $\Phi$ we then have the following situation. If the temperature is smaller (bigger) than the critical one, $T<T_c$ ($T>T_c$), the dominant saddle point is HFS (FSC). Therefore, we reach a similar conclusion as in three dimensions: flat space undergoes a phase transition if it is put at some electric potential and gets heated up sufficiently.

\subsection{Thermodynamics of flat space Ach\'ucarro--Ortiz}\label{se:2.3}

Starting from the free energy \eqref{eq:phase10} all thermodynamical variables of interest can be derived. We start with entropy.
\eq{
S = -\frac{\partial Y}{\partial T}\Big|_{\Phi} = 2\pi X_h
}{eq:phase13}
Thus, the entropy is essentially the dilaton evaluated at the horizon, as expected from previous results \cite{Gegenberg:1995pv,Kunstatter:1998my,Davis:2004xi,Grumiller:2007ju} and from comparison with Wald's entropy \cite{Wald:1993nt,Iyer:1994ys}.

The conjugate variable to the electric potential should be the charge $q$. Consistently with this expectation we find
\eq{
q = -\frac{\partial Y}{\partial \Phi}\Big|_T 
}{eq:phase14}
which explicitly evaluates to $q= -4\pi^2\,T^2/\Phi^3$.
In the 3d picture the charge $q$ corresponds to the angular momentum of the FSC solution.

As it must be, the first law of thermodynamics holds.
\eq{
\extd Y = -S\,\extd T - q\,\extd\Phi
}{eq:phase16}
While this is merely a consistency check, we stress at this point that none of the results above could have been obtained without the last boundary term in the action \eqref{reducedaction}, which contributes essentially to the free energy $Y$. (By contrast, the first boundary term is irrelevant and can be dropped.)

The specific heat at constant electric potential turns out to be positive.
\eq{
C_\Phi = T \,\frac{\partial S}{\partial T}\Big|_\Phi = S = 4\pi^2\,\frac{T}{\Phi^2}
}{eq:phase15}
The last equality shows that entropy and specific heat are compatible with the third law, in the sense that both vanish as $T$ tends to zero. In fact, specific heat tends to zero linearly, exactly like a free Fermi gas at low temperature. Also this result was found already in the 3d discussion \cite{Bagchi:2013lma}.

The electric susceptibility is positive as well, which shows that 2d FSCs have the same property as standard electric materials.
\eq{
\chi = \frac{\partial q}{\partial\Phi}\Big|_T = \frac{3}{4\pi}\,S^2 = 12\pi^2 \,\frac{T^2}{\Phi^4}
}{eq:phase43}
Moreover, again we get compatibility with the third law.

To conclude this section, we have succeeded in providing an intrinsic 2d formulation of FSCs and their thermodynamics. We have found that the phase transition discovered in three dimensions persists also in two dimensions, i.e., for a specific Einstein--Maxwell--Dilaton model. To facilitate the comparison between 2d and 3d results we collect a dictionary in table \ref{tab:1}.

It is now interesting to ask whether there are also other 2d Einstein--Maxwell--Dilaton models exhibiting the features we found above. We address this question in the next section, where it will be answered in the affirmative.

\renewcommand{\thefigure}{F.\arabic{figure}}
\renewcommand{\thetable}{T.\arabic{table}}
\begin{center}
\begin{table}
\centering
\hspace*{-0.1truecm}
\begin{tabular}{|l|l|l|}
 \hline
& 2d & 3d \\ \hline
Einstein & metric $g_{ab}$ & constant $\varphi$-slice \\
Maxwell & gauge field $A_a$ & $\varphi$-twist in $\tau$-identification \\
Dilaton & dilaton field $X$ & radial coordinate $r$ \\ 
Action & Eq.~\eqref{reducedaction} & Eq.~\eqref{flatgamma3} with $\alpha=\tfrac12$ \\  \hline
Boundary term & normal derivative of dilaton & 1/2 GHY \\
Boundary conditions & Eqs.~\eqref{eq:bc} & see \cite{Barnich:2006av,Bagchi:2012yk} \\
EOM & Eqs.~\eqref{EOM1}-\eqref{EOM3} & $R_{AB} = 0$ \\ 
Solutions & Eqs.~\eqref{eq:phase5}-\eqref{eq:phase7} & Eq.~\eqref{FSC3} \\ \hline
1$^{\rm st}$ constant of motion & $M$ (2d mass, $M>0$) & $r_+^2$ (twice 3d mass)\\
2$^{\rm nd}$ constant of motion  & charge $q$ & angular momentum $J=-r_+r_0$\\
1$^{\rm st}$ conjugate quantity & temperature $T=\frac{q^2}{2\pi X_h^3}$ & temperature  $T=\frac{r_+^2}{2\pi r_0}$ \\
2$^{\rm nd}$ conjugate quantity & electric potential $\Phi$ & angular potential $\Omega$ \\ \hline
Solutions with horizon & 2d FSC & 3d FSC \\
Solutions without horizon & HFS with electric potential & HFS with angular potential \\
Phase transition & between 2d FSC and HFS & between 3d FSC and HFS \\
Critical temperature & $T_c=\frac{\Phi}{2\pi}$ &  $T_c=\frac{\Omega}{2\pi}$ \\ \hline
Free energy (FSC) & $Y(T,\,\Phi) = -2\pi^2 \,\frac{T^2}{\Phi^2}$ & $F(T,\,\Omega) = -2\pi^2 \,\frac{T^2}{\Omega^2}$\\
Entropy (FSC) & $S = 2\pi X_h = 2\pi\,\frac{|q|}{\sqrt{M}}$ & $S = A = 2\pi |r_0|$ (note: $G_3=\tfrac14$)\\
First law (FSC) & $\extd Y = - S\,\extd T - q\,\extd\Phi$ &  $\extd F = - S\,\extd T - J\,\extd\Omega$ \\ 
Specific heat (FSC) & $C_\Phi = S = 4\pi^2 \frac{T}{\Phi^2}$ & $C_\Omega = S = 4\pi^2 \frac{T}{\Omega^2}$\\
\hline
\end{tabular}
\caption[Dictionary between two and three dimensions]{Dictionary between 2d and 3d}
 \label{tab:1}
\end{table}
\end{center}

\section{Asymptotically mass-dominated dilaton gravity}\label{se:3}

In this section we generalize the results of the previous section to a specific class of 2d dilaton gravity models, namely those models whose asymptotic behavior is dominated by the mass term in the line-element. Interestingly, it is precisely this class of models that so far had been neglected in discussions of 2d dilaton gravity thermodynamics.

In section \ref{se:3.0} we define the notion of asymptotic mass-domination and show that this case is not covered by results in the literature.
In section \ref{se:3.1} we provide the boundary terms required for a well-defined variational principle. 
In section \ref{se:3.2} we calculate the Euclidean partition function and show the persistence of the phase transition, given some assumptions that we spell out.
In section \ref{se:3.3} we conclude with a discussion of thermodynamical properties.

\subsection{Asymptotic mass-domination and absence of confinement}\label{se:3.0}

Einstein--Maxwell--Dilaton gravity in two dimensions is defined by the bulk action (see \cite{Grumiller:2002nm} for a review on 2d dilaton gravity and \cite{Strobl:1994eu,Louis-Martinez:1995rq} for some early papers on Einstein--Maxwell--Dilaton gravity)
\begin{equation}
  I_{\textrm{\tiny bulk}} =
    - \frac{1}{2} \int_{\mathrlap{\mathcal M}} \; \extd^2 x \sqrt{g} \,\left( XR - U(X) (\partial X)^2 - 2 V(X) \right)
    + \int_{\mathrlap{\mathcal M}} \; \extd^2 x \sqrt{g}\, f(X) F^{ab} F_{ab} \, . \label{dilatonbulk}
\end{equation}
The kinetic potential $U(X)$, the dilaton self-interaction potential $V(X)$, and the strength of the coupling function $f(X)$ of the Maxwell field define the model.

The action \eqref{dilatonbulk} yields the EOM
\begin{subequations}
\label{eq:eom}
\begin{align}
  \mathcal{E}_{ab}=\nabla_a \partial_b X - g_{ab} \nabla^2X  + U(X) (\partial_a X) (\partial_b X)- \tfrac{1}{2}g_{ab} U(X) (\nabla X)^2  & \nonumber \\
   - g_{ab} V(X) + 4f(X)\big( F_a{}^c F_{bc} - \tfrac14\,g_{ab}F^{cd} F_{cd}\big) &= 0 \label{EOM1a} \\
  \mathcal{E}_X=R + 2 U(X) \nabla^2 X + U^\prime(X) (\partial X)^2 - 2 V^\prime(X) - 2 f^{\prime}(X) F^{ab} F_{ab} &= 0 \label{EOM2a} \\
  \mathcal{E}^{b}=\nabla_{a} ( f(X) F^{ab} ) &= 0 \,. \label{EOM3a}
\end{align}
\end{subequations}

Generic solutions of the EOM are parametrized by two constants of motion, mass $M$ and $U(1)$ charge $q$, and can be written in the form
\begin{subequations}
\label{eq:sols}
\begin{align}
  \extd s^2 = \xi(r)\,\extd \tau^2+\frac{\extd r^2}{\xi(r)} & \qquad \partial_rX(r)=e^{-Q(X)} \label{eq:sol1} \\
  F_{ab} = \frac{q}{4f(X)}\, \epsilon_{ab}\, . \label{eq:sol2}
\end{align}
The ``Killing norm'' $\xi$ as a function of the dilaton field is given by
\begin{equation}
  \xi(X) = e^{Q(X)}\,\Big(M+w(X)+\frac{q^2}{4}h(X)\Big) \, , \label{Killingnorm}
\end{equation}
where the functions $Q(X)$, $w(X)$ and $h(X)$ are defined as
\begin{align}
  Q(X) &:= Q_0 + \int \! \extd X \, U(X) \label{eq:Q} \\
  w(X) &:= w_0 - 2 \int \! \extd X \, V(X) e^{Q(X)} \label{eq:w} \\
  h(X) &:= \int \! \extd X \, \frac{1}{f(X)} e^{Q(X)} \label{eq:h}
\end{align}
\end{subequations}
By an appropriate choice of $w_0$ the constant of motion $M$ can be restricted to a convenient range. The constant $Q_0$ can be made to vanish by a redefinition of the coordinates.

The solutions \eqref{eq:sols} exhibit a Killing vector field $\partial_{\tau}$ with norm $\xi(X)$ (which justifies the name ``Killing norm''). The constant $X$ hypersurfaces on which the Killing norm vanishes,  $\xi(X)=0$, represent Killing horizons in the Minkowski analogue of the theory. We denote the value of the dilaton field at the outermost Killing horizon by $X_h$. It is determined as the largest (real and positive) solution $X_h$ of the equation
\eq{
M+w(X_h)+\frac{q^2}{4}\,h(X_h)=0\,.
}{eq:phase45}

To determine the gauge field again we choose axial gauge.
\begin{equation}
   A_{\tau} = -\frac{q}{4}\left( h(X)-h(X_h)\right)+A_{\tau}(X_h)\qquad A_r = 0 
\label{eq:A}
\end{equation}
The remaining gauge freedom is in the constant $A_{\tau}(X_h)$, which determines the gauge potential on the horizon. The proper electric potential between some conducting cavity wall at $X=X_c$ and the horizon $X=X_h$ is given by
\begin{equation}
  \Phi(X_c)=A_{\tau}(X_c)-A_{\tau}(X_h)\,.
\end{equation}

The Hawking temperature of the 2d spacetime is obtained from the standard Euclidean field theory argument:
In order to avoid conical singularities at the horizon, the Euclidean time is assumed to be periodic $\tau \sim \tau + \beta$, with
\begin{equation}
  T=\beta^{-1}=\frac{\partial_r\xi}{4\pi}\bigg|_{r_h} = \frac{w^\prime(X_h)+\frac{q^2}{4}\,h^\prime(X_h)}{4\pi}\,. \label{eq:beta}
\end{equation}

Different classes of models can be characterized by the asymptotic behavior of the functions $w(X)$ and $h(X)$ in the limit $X\rightarrow \infty$.\footnote{%
Two comments are in order. First, there is also a condition on the function $U(X)$ in all cases, namely  $\lim_{X\to\infty} U(X) > -\frac{2}{X}$. This inequality ensures that the asymptotic region is reached as $X\to\infty$. However, if one demands in addition that the asymptotic region $X\to \infty$ corresponds to the coordinate $r\to\infty$, one needs the stronger condition $\lim_{X\to\infty} U(X)\ge-\frac{1}{X}$. This is what we assume in the following. Second, below we list limiting cases that tend to $\infty$ or $0$ but not cases where the quantities tend to finite values. This is so, because the case of finite limits is equivalent to the case of vanishing limits, upon suitable redefinitions of constants of motion.
} The possibilities studied so far in the literature are enumerated below.
\begin{enumerate}
 \item {\bf Asymptotic dilaton domination.} This case is defined by the properties
\eq{
\lim_{X\to\infty}w(X) = +\infty\qquad \lim_{X\to\infty}h(X) = 0\,.
}{eq:phase20}
We choose the characterization ``asymptotic dilaton domination'' because the dilaton-dependent term $w(X)$ dominates asymptotically over the mass term $M$ in the Killing norm \eqref{Killingnorm}. Thermodynamics for this case was studied extensively in \cite{Grumiller:2007ju}.
 \item {\bf Asymptotic dilaton domination (with confining $\boldsymbol{U(1)}$ charge).} This case is defined by the properties
\eq{
\lim_{X\to\infty}w(X) = +\infty\qquad \lim_{X\to\infty} |h(X)| = \infty \,.
}{eq:phase21}
The most dominant term in the Killing norm \eqref{Killingnorm} is either of the dilaton-dependent terms (confinement requires additionally $|f(X)V(X)|<\infty$, in which case the $h(X)$-term is the most dominant term in the Killing norm).
Thermodynamics for the confining case was studied recently in \cite{Grumiller:2014oha}, where it was also explained why the attribute ``confining'' is justified. The non-confining case works analog to the first case above.
 \item {\bf Asymptotic confinement domination.} This case is defined by the properties
\eq{
\lim_{X\to\infty}w(X) = 0\qquad \lim_{X\to\infty} |h(X)| = \infty \,.
}{eq:phase22}
The most dominant term in the Killing norm \eqref{Killingnorm} is the one proportional to $h(X)$, which comes from the confining $U(1)$ charge. Thermodynamics for this case was also studied recently in \cite{Grumiller:2014oha}.
\end{enumerate}

\paragraph{Definition of asymptotic mass-domination.}
We say that a model exhibits the property of asymptotic mass-domination if the associated functions $w(X)$ and $h(X)$ both tend to zero asymptotically.
\eq{
\boxed{
\phantom{\Big(}
\lim_{X\to\infty}w(X) = 0\qquad \lim_{X\to\infty} h(X) = 0
\phantom{\Big(}
}
}{eq:phase23}
In this case the most dominant term in the Killing norm \eqref{Killingnorm} at large values of $X$ is the mass term $M$. From the definition \eqref{eq:phase22} it is evident that asymptotic mass-domination implies that the $U(1)$ charge cannot be confining. 
For the Penrose diagram of asymptotically mass-dominated geometries we refer to appendix \ref{app:A}.

A consequence of asymptotic mass-domination is that curvature 
\eq{
R = -e^{-Q}\,\Big(w''+\frac{q^2}{4}\,h''+U\big(w'+\frac{q^2}{4}\,h'\big)+U'\big(M+w+\frac{q^2}{4}\,h\big)\Big)
}{eq:phase34}
tends to zero asymptotically. This is so, because the pre-factor $e^{-Q}$ cannot grow as strong as (or stronger than) $X^2$ due to the inequality $U>-2/X$. This implies that the combination $e^{-Q} U' M$ tends to zero at large $X$, and all other terms are even smaller due to asymptotic mass-domination. Therefore, asymptotic mass-domination leads to spacetimes that are asymptotically flat (though not necessarily with the same asymptotic behavior as in section \ref{se:2}). We call generic solutions of this type ``generalized flat space cosmologies'' (generalized FSCs).

Checking the comprehensive list of well-studied 2d dilaton gravity models in Table 1 of \cite{Grumiller:2006rc}, with the (rather formal) exception of spherical reduction from less than three dimensions none of these models is asymptotically mass-dominated. Thus, asymptotic mass-domination appears to be a rare property.
Looking back at the flat space Ach\'ucarro--Ortiz model \eqref{reducedaction} we obtain the potentials
\eq{
U(X)=V(X)=Q(X)=w(X)=0\qquad f(X) = \frac18\,X^3\qquad h(X)=-\frac{4}{X^2}\,.
}{eq:AOflat}
The flat space Ach\'ucarro--Ortiz model is compatible with the limits \eqref{eq:phase23}, and thus provides the first example of Einstein--Maxwell--Dilaton models with asymptotic mass-domination. 

\subsection{Variational principle}\label{se:3.1}

We consider now a large class of Einstein--Maxwell--Dilaton models with asymptotic mass-domination, by assuming
\begin{align}
  U(X) = -\frac{a}{X}+o(X^{-1}),\,\, a\le1 \qquad
  V(X) = o(1) \qquad
  \frac{1}{f(X)} = o(1) \,. \label{conditions}
\end{align}
Additionally we assume that all functions are asymptotically a monomial in $X$, in order to avoid subtleties with essential asymptotic singularities.

In order to set Dirichlet boundary conditions on the metric a term analogous to the GHY term has to be added to the bulk action \eqref{dilatonbulk}.
\begin{equation}
  I_{\textrm{\tiny GHY}} = - \int_{\mathrlap{\partial \mathcal M}} \; \extd x \sqrt{\gamma}\, X K 
\end{equation}
However, here we allow for a more general action and do not restrict ourselves to Dirichlet boundary conditions. The GHY-term is therefore added to the action $I_{\textrm{\tiny bulk}}$ with an arbitrary coefficient $\alpha$. We are going to show that this action
\begin{equation}
  I = I_{\textrm{\tiny bulk}} + \alpha I_{\textrm{\tiny GHY}} \label{action1}
\end{equation}
does not yield a well-defined variational principle in the above sense for any value of $\alpha$.

The first variation of the action \eqref{action1} reads
\begin{multline}
  \delta I =
    - \frac{1}{2} \int_{\mathrlap{\mathcal M}} \; \extd^2 x \sqrt{g} \; \mathcal{E}^{ab}_g \; \delta g_{ab}
    - \frac{1}{2} \int_{\mathrlap{\mathcal M}} \; \extd^2 x \sqrt{g} \; \mathcal{E}_X \; \delta X
    + \int_{\mathrlap{\mathcal M}} \; \extd^2 x \sqrt{g} \; \mathcal{E}^{a}_A \; \delta A_a
    + \\ 
    + \frac{1}{2} \int_{\mathrlap{\mathcal \partial M}} \; \extd x \sqrt{\gamma}\, \Big(
      \big(
        (1 - \alpha) X K (2\gamma^{ab}-g^{ab})
        + 2 (1 - \alpha) n^a \gamma^{bc} \partial_c X
        - \gamma^{ab} n^c \partial_c X
      \big) \delta g_{ab}
      +  \\ 
      + (1 - \alpha) \gamma^{ab} X n^c \partial_c \delta g_{ab}
    \Big)
    + \int_{\mathrlap{\mathcal \partial M}} \; \extd x \sqrt{\gamma} \,\left( U n^a \partial_a X
    - \alpha K \right) \delta X
    - 4\int_{\mathrlap{\mathcal \partial M}} \; \extd x \sqrt{\gamma}\,n_a F^{ab}\delta A_b
\label{eq:whyhasthisnolabel}
\end{multline}
The boundary conditions
\begin{subequations}
 \label{bcs}
\begin{align}
 g_{ab}  &= \begin{pmatrix}
       g_{\tau\tau} = e^{Q}M \,[1 + o(1)] & g_{\tau r} = o(1) \\
       & g_{rr} = e^{-Q}\,\frac1M\,[1 + o(1)]
      \end{pmatrix} \label{eq:bc1a} \\
 A_a &= \begin{pmatrix}
       A_\tau = o(1) \\
       A_r = o(1)
      \end{pmatrix} \label{eq:bc2a} \\
a< 1: X & = [(1-a)r]^{\frac{1}{1-a}}\,[1 + o(1)] \qquad  a=1: X=e^r\,[1 + o(1)] \label{eq:bc3a}
\end{align}
\end{subequations}
follow from the conditions \eqref{conditions} and the EOM \eqref{eq:eom}. 
They are generalizations of the  Euclidean flat space boundary conditions \eqref{eq:bc}.
For simplicity we choose the gauge $g_{rr}g_{\tau\tau}=1$, although this condition can be weakened. Moreover, we use again axial gauge $A_r=0$. If we now set $\alpha=1$ then only the last term in the second line of \eqref{eq:whyhasthisnolabel} gives a finite contribution to the first variation of the action. For any other value of $\alpha$ there are even more non-vanishing terms, which do not cancel each other. The variational principle is therefore not well-defined for any value of $\alpha$, though $\alpha=1$ almost works, in the sense that there is only one offending term remaining. 

This remaining term can be canceled by adding to the action a suitable boundary counterterm. Inspired by the last boundary term in the flat space Ach\'ucarro--Ortiz action \eqref{reducedaction} we propose the boundary counterterm
\begin{equation}
  I_{\textrm{ct}} = \frac{1}{2} \int_{\mathrlap{\partial \mathcal M}} \; \extd x \sqrt{\gamma} \,n^a \partial_a X \, .
\end{equation}
Varying this term indeed precisely cancels the finite contribution in the variation of the action \eqref{eq:whyhasthisnolabel} for $\alpha=1$ or when $U$ falls off faster than $1/X$.

Thus, the full action
\begin{equation}
  \Gamma = I_{\textrm{\tiny bulk}} + \alpha I_{\textrm{\tiny GHY}}+I_{\textrm{ct}} \qquad \alpha=1  \label{fullaction}
\end{equation}
has a well-defined variational principle, $\delta \Gamma=0$, for the generalized flat space boundary conditions \eqref{bcs}. To avoid confusion we note that the value of $\alpha$ is irrelevant for models with $U(X)=o(1/X)$, including the special case $U(X)=0$. This is why in the flat space Ach\'ucarro--Ortiz model the value of $\alpha$ was irrelevant (in fact, we chose it as $\alpha=\tfrac12$, as this was the value we obtained from Kaluza--Klein reduction, but effectively this term is zero there).

\subsection{General phase transitions in two dimensions}\label{se:3.2}

Again there are two Euclidean saddle points that contribute for any given values of temperature $T$ and electric potential $\Phi$. The contribution from HFS \eqref{eq:phase17} is the same as before and leads again to the free energy \eqref{eq:phase18}. 

The free energy of generalized FSCs, \eqref{eq:sols} with \eqref{eq:phase23}, reads
\eq{
\boxed{
\phantom{\Big(}
 Y(T,\,\Phi) = - \tfrac12\,M - 2\pi T X_h - q\Phi 
\phantom{\Big(}
}
}{eq:phase24}
where $\Phi$ is again defined as
\begin{equation}
\Phi = \frac{q}{4}\,h_h\, .
\end{equation}
Here and in what follows subscripts $h$ imply that the corresponding quantity is evaluated at the horizon, e.g.~$h_h=h(X_h)$.
Expressing the free energy $Y$ in terms of quantities evaluated at the horizon yields
\eq{
Y=\frac12\,\big(w_h-X_h w_h^\prime\big)-\frac{q^2}{8}\,\big(h_h+X_hh_h^\prime\big)\,.
}{eq:phase50}
The quantities $w_h$ and $h_h$ are related to mass $M$ by the horizon condition \eqref{eq:phase45}, and their derivatives to temperature $T$ by \eqref{eq:beta}.

If the free energy \eqref{eq:phase24} crosses the line $Y(T,\,\Phi)=-\tfrac12$ for a 1-parameter family of combinations of temperature and electric potential, then there will be a phase transition along this critical line, precisely of the same nature as in section \ref{se:2}. 

We argue now that such phase transitions must always exist, provided our assumptions hold, the functions are sufficiently smooth and obey suitable conditions that we derive below. Suppose we find some solution whose free energy is smaller (larger) than $-\tfrac12$. Then generically by smoothness there must be an open region in the parameter space spanned by $M$ and $q$ where the free energy is smaller (larger) than $-\tfrac12$. Moreover, almost any solution must belong to either of these two ``basins'', $Y<-\tfrac12$ or $Y>-\tfrac12$. In addition, by continuity there must be a co-dimension 1 surface in the parameter space where $Y=-\tfrac12$, which separates the two basins, provided both are non-empty. This co-dimension 1 surface is the critical surface where the phase transition takes place between HFS and some generalized FSC solution.

In order for the argument above to work, however, we still have to prove that both basins are non-empty. Indeed, a simple way to avoid a phase transition would be to find a model which has $Y\geq 0$ in the whole parameter space, so that HFS would always be the dominant saddle point. 

We restrict ourselves first to the special case $w=0$. In that case the free energy \eqref{eq:phase50} simplifies to
\eq{
Y = -\frac{q^2}{8}\,\big(h_h + X_h h_h^\prime\big) \,.
}{eq:phase28}
There are now two possibilities. If $Y\geq 0$ for all possible values of $(M,q)$ then there can never be a phase transition. If $Y<0$ for some values of $(M,q)$ there could be a phase transition, so a necessary condition that the function $h$ must fulfill is
\eq{
h_h + X_h h_h^\prime > 0\,.
}{eq:phase47}
Note that the second term must be positive so that Hawking temperature \eqref{eq:beta} is positive, while the first term is usually negative due to asymptotic mass-domination, and in particular it must be negative if mass is required to be positive. Thus, the inequality does not hold automatically.
Assuming that the inequality \eqref{eq:phase47} holds [at least for some choices of $(M,q)$], the key observation is now that for any solution $(M,q)$ of the horizon condition 
\eq{
M=-\frac{q^2}{4}\,h_h
}{eq:phase26}
we can find another solution by rescaling $(M,q)\to(\lambda^2M,\,\lambda q)$, which results in $Y\to \lambda^2 Y$. Since $Y<0$ we have therefore necessarily an open set of solutions in either of the two basins, $Y<-\tfrac12$ and $Y>-\tfrac12$, which concludes our argument.

If $w\neq 0$ the discussion is a bit more involved, though it is very simple to verify the existence of a phase transition on a case-by-case basis. Here we argue that the existence of non-vanishing $w$ typically helps to promote the free energy $Y$ to negative values, in which case arguments similar to the one in the previous paragraph can be used. Namely, consider the limit of very small charges, so that all terms with $h_h$ and $h_h^\prime$ can be neglected. In that case free energy \eqref{eq:phase50} simplifies to
\eq{
Y = \frac12\,\big(w_h-X_hw_h^\prime\big)+{\cal O}(q^2)\,.
}{eq:phase51}
Positivity of the Hawking temperature \eqref{eq:beta} again requires the second term to be positive, but asymptotic mass-domination now also implies usually\footnote{%
For monotonic functions $w$ this is certainly true. Even for non-monotonic functions it still must be true if we demand positivity of the mass $M$.
} that the first term is negative. But then $Y$ is always negative, so the case $Y\geq 0$ cannot arise in the small-charge limit. For finite charges $q$ the terms in \eqref{eq:phase51} can provide the existence of a phase-transition even if the inequality \eqref{eq:phase47} is violated.

Thus, we have shown that the phase transition between HFS and generalized FSCs occurs fairly generically. For the simple class of models with $w=0$ it always occurs, provided the inequality \eqref{eq:phase47} holds. This is true in particular for any monomial function $h=-a^2/X^{n-1}$ with $a\in\mathbb{R}$ and $n>2$.

\subsection{Thermodynamics of asymptotically mass-dominated dilaton gravity}\label{se:3.3}

\subsubsection[General remarks on the ensemble \texorpdfstring{$Y(T,\,\Phi)$}{Y(T,Phi)}]{General remarks on the ensemble $\boldsymbol{Y(T,\,\Phi)}$}\label{se:3.3.1}

In the previous subsection we showed that the phase transition between HFS and generalized FSCs persists, given some conditions on the functions $w$ and $h$. Here we address other thermodynamical aspects.

From the free energy \eqref{eq:phase24} it is straightforward to derive again all thermodynamical quantities of interest. Entropy \eqref{eq:phase13} and charge \eqref{eq:phase14} yield the same results as before, and also the first law \eqref{eq:phase16} still holds, as it must. The only quantities that change are specific heat and electric susceptibility. As we shall see they are not necessarily positive.

\subsubsection{Specific heat at constant electric potential}

Introducing the abbreviation
\eq{
\tilde w = w + \frac{q^2}{4}\,h
}{eq:phase29}
we obtain
\begin{equation}
C_{\Phi}=2\pi \frac{\tilde{w}^{\prime}_h}{\tilde{w}^{\prime\prime}_h-\frac{q^2}{2}\frac{(h_h^{\prime})^2}{h_h}}\, .
\end{equation}
Since the numerator must be positive in order to have positive temperature \eqref{eq:beta}, the requirement of positive specific heat leads to the inequality
\begin{equation}
\tilde{w}^{\prime\prime}_h>\frac{q^2}{2}\frac{(h_h^{\prime})^2}{h_h}\,. 
\label{eq:phase31}
\end{equation}
In the special case $w=0$ it simplifies to the inequality
\begin{equation}
h^{\prime\prime}_h > 2\frac{(h^{\prime}_h)^2}{h_h}\,. \label{hinequality}
\end{equation}

A simple example of an asymptotically mass-dominated model that fails to obey the inequalities \eqref{eq:phase31} and \eqref{hinequality} is given by
\eq{
U(X)=V(X)=Q(X)=w(X)=0\qquad f(X) = \tfrac12\,X^{3/2}\qquad h(X) = -\frac{4}{\sqrt{X}}\,.
}{eq:phase30}
While this model is not motivated by some particular physical question, it shows that positivity of specific heat is not automatic, but has to be checked on a case-by-case basis.

\subsubsection[Alternative ensemble \texorpdfstring{$F(T,\,q)$}{F(T,q)}]{Alternative ensemble $\boldsymbol{F(T,\,q)}$}

The Helmholtz free energy $F(T,\, q)$ 
\begin{equation}
F(T,\,q) = - \tfrac12\, M - 2 \pi T X_h \label{freeenergy}
\end{equation}
is obtained by a Legendre transformation of the free energy $Y(T,\,\Phi)$ from \eqref{eq:phase24}. One can derive the Helmholtz free energy directly from an on-shell action by adding to \eqref{fullaction} another boundary term.
\begin{equation}
I_{FA}=4\int_{\partial \mathcal{M}}\!\!\!\!\!\!\extd x\sqrt{\gamma}\,n_{a}A_{b}F^{ab}
\end{equation}
However, then we would need to impose Neumann boundary conditions instead of Dirichlet boundary conditions on the gauge field $A_{a}$. The reason why we wanted to impose Dirichlet boundary conditions on the gauge field in the flat-space Ach\'ucarro--Ortiz model was motivated by our Kaluza--Klein reduction, but we can now give an intrinsic 2d argument why the ensemble $Y(T,\,\Phi)$ is preferred over the ensemble $F(T,\,q)$.

Namely, consider specific heat at constant charge $C_q$. 
\begin{equation}
C_q=T\,\frac{\partial S}{\partial T}\Big|_q=2\pi\, \frac{\tilde{w}^{\prime}_h}{\tilde{w}^{\prime\prime}_h}
\end{equation}
This is the same result as previously obtained in \cite{Grumiller:2007ju}. In order to have positive temperature we require again $\tilde{w}^\prime_h > 0$. However, if both functions $w$ and $h$ are monomial in $X$ and vanish in the asymptotic limit, then we have $\tilde{w}_h^{\prime \prime} < 0$. Thus, the specific heat at constant charge is negative and the respective ensemble is not well-defined, as the Gaussian fluctuations \eqref{eq:angelinajolie} destabilize system. This is the intrinsic 2d reason why we do not use the free energy \eqref{freeenergy} as thermodynamic potential. 

\subsubsection{Electric susceptibility}

Electric susceptibility can be expressed in terms of specific heats as
\eq{
\chi = \frac{\partial q}{\partial\Phi}\Big|_T = \frac{4C_\Phi}{h_h C_q}\,.
}{eq:phase52}
As discussed before both $h_h$ and $C_q$ are usually negative for asymptotically mass-dominated models, so that the sign of electric susceptibility equals the sign of specific heat at constant electric potential. 

\subsubsection{Simple family of examples}

We provide now a simple explicit family of models.
\eq{
U(X)=V(X)=Q(X)=w(X)=0\qquad f(X)=\frac{X^n}{4(n-1)}\qquad h(X) = -\frac{4}{X^{n-1}}
}{eq:phase25}
In order to achieve asymptotic mass-domination we demand $n>1$. The flat space Ach\'ucarro--Ortiz model \eqref{eq:AOflat} is recovered for $n=3$. 

We are thus led to the following set of solutions
\begin{equation}
\xi(X)=M-\frac{q^2}{X^{n-1}} \qquad X(r)=r \qquad F_{ab}=\frac{q(n-1)}{X^{n}}\epsilon_{ab}\, .
\end{equation}
The corresponding Penrose diagram is discussed in appendix \ref{app:A}.
Temperature and electric potential of these generalized 2d FSCs are given by
\begin{equation}
T=\frac{n-1}{4\pi}q^2X_h^{-n} \qquad \Phi=-qX_h^{1-n}\, ,
\end{equation}
where $X_h$ in terms of the conserved quantities  $M$ and $q$ is given by
\begin{equation}
X_h=\left(\frac{q^2}{M}\right)^{\frac{1}{n-1}}\, .
\end{equation}
It is useful to express the two parameters $M$, $q$ in terms of $T$ and $\Phi$.
\eq{
M =\frac{4\pi}{n-1}T\left(\frac{4\pi}{n-1}\frac{T}{\Phi^2}\right)^{\frac{1}{n-2}}\qquad
q =-\frac{4\pi}{n-1}\frac{T}{\Phi}\left(\frac{4\pi}{n-1}\frac{T}{\Phi^2}\right)^{\frac{1}{n-2}}
}{eq:phase48}

The thermodynamic potential $Y(T,\Phi)$ is given by
\begin{equation}
Y(T,\Phi)= -\big(\tfrac n2-1\big)\,M = -\big(\tfrac n2-1\big)\,\frac{4\pi}{n-1} \,T\,\left(\frac{4\pi}{n-1}\frac{T}{\Phi^2}\right)^{\frac{1}{n-2}}\, .\label{gFSCY}
\end{equation}
Entropy, charge and first law follow again the general discussion in section \ref{se:3.3.1}. 

The special case $n=2$ leads to vanishing free energy, $Y=0$, and provides the simplest example for an asymptotically mass-dominated model without phase transition.
For $n>2$ the inequality \eqref{eq:phase47} holds and the phase transition between HFS and generalized FSCs occurs at the critical temperature
\begin{equation}
T_c^{n-1}=\frac{n-1}{4\pi}\,\Phi^2\left(\frac{n-1}{n-2}\frac{1}{4\pi}\right)^{n-2}\,.
\end{equation}

The specific heat at constant electric potential is given by
\begin{equation}
C_{\Phi}=\frac{S}{n-2}\, .
\end{equation}
Thus, specific heat is positive only for $n>2$. This is compatible with the inequality \eqref{hinequality}.
Similarly, electric susceptibility is positive for $n>2$.
\eq{
\chi = \frac{n}{n-2}\,\Big(\frac{S}{2\pi}\Big)^{n-1}
}{eq:phase46}
Thus, for the simple family of models \eqref{eq:phase25} the inequality \eqref{eq:phase47} that guarantees the existence of the phase transition simultaneously guarantees the positivity of specific heat and electric susceptibility.


\section*{Acknowledgments}

We are grateful to St\'ephane Detournay, Robert McNees and Joan Sim\'on for pleasant and fruitful collaborations on related topics.

AB was supported by an INSPIRE award of the Department of Science and Technology, India and by the project M~1508 of the Austrian Science Fund (FWF). DG, JS and FS were supported by the START project Y~435-N16 of the FWF and the FWF projects I~952-N16, I~1030-N27 and P~27182-N27.

\begin{appendix}

\section{Penrose diagram}\label{app:A}

In this appendix we discuss global properties of solutions for 2d dilaton gravity models with asymptotic mass-domination. We follow the general construction by Kl\"osch and Strobl \cite{Klosch:1996fi} 
and assume Minkowski signature in the whole appendix.

For simplicity we assume that the functions $w$ and $h$ are chosen such that there is only one Killing horizon, though the whole discussion in the main text also would work for multiple Killing horizons (the outermost Killing horizon would then play the role of the Killing horizon used in the main text).

\begin{figure}
\centering
\begin{tikzpicture}
\node (I)   at ( 0, 2) {I};
\node (II)  at ( 0,-2) {II};
\node (III) at (-1, 0) {III};
\node (IV)  at ( 1, 0) {IV};

\path (I) +( 90:2) coordinate[label=90:$i^+$] (Itop)
          +(-90:2) coordinate (Ibot)
          +(  0:2) coordinate (Iright)
          +(180:2) coordinate (Ileft);

\path (II) +( 90:2) coordinate (IItop)
           +(-90:2) coordinate[label=-90:$i^{-}$] (IIbot)
           +(180:2) coordinate (IIleft)
           +(  0:2) coordinate (IIright);

\draw [thick,decoration={snake,segment length=0.343cm}] (Ileft)
  --(Itop)   node[midway,above left]  {$\cal{J}^+$}
  --(Iright) node[midway,above right] {$\cal{J}^{+}$}
  decorate { --(IIright) }
  --(IIbot)  node[midway,below right] {$\cal{J}^{-}$}
  --(IIleft) node[midway,below left]  {$\cal{J}^{-}$}
  decorate { --cycle };

\draw [very thin,dashed]
  (Ileft) --(IIright) node[near start,below left]  {$\cal{H}^+$}
                      node[near end,above right]   {$\cal{H}^-$}
  (Iright) --(IIleft) node[near start,below right] {$\cal{H}^+$}
                      node[near end,above left]    {$\cal{H}^-$};

\draw [very thin] (Ileft) --(Iright);
\draw [very thin] (Ileft) to [out=20,in=160] (Iright);
\draw [very thin] (Ileft) to [out=40,in=140] (Iright);
\draw [very thin] (Ileft) to [out=-20,in=-160] (Iright);
\draw [very thin] (Ileft) to [out=-40,in=-140] (Iright);

\draw [very thin] (IIleft) --(IIright);
\draw [very thin] (IIleft) to [out=20,in=160] (IIright);
\draw [very thin] (IIleft) to [out=40,in=140] (IIright);
\draw [very thin] (IIleft) to [out=-20,in=-160] (IIright);
\draw [very thin] (IIleft) to [out=-40,in=-140] (IIright);
\end{tikzpicture}
\caption[Penrose diagram of two-dimensional flat space cosmologies]{Penrose diagram of 2d FSCs}
\label{fig:penrose_diagram}
\end{figure}
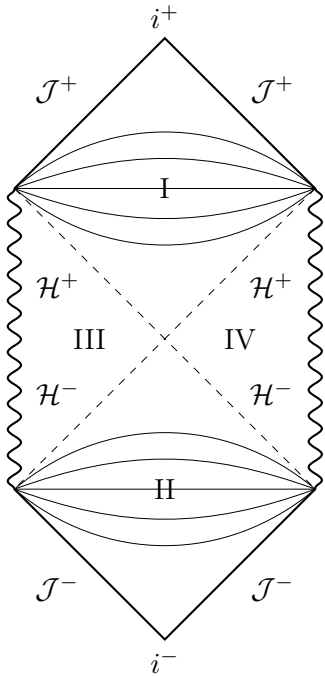

For simplicity we further assume that $w=0$, though again this assumption can be relaxed without yielding anything novel. Moreover, without any essential loss we further assume that $h$ is a monomial in $X$ (if it is a generalized polynomial in $X^{-\alpha}$ we just take the asymptotically most dominant monomial contribution, which is sufficient for discussing the asymptotic boundary of the Penrose diagram).
\eq{
h(X) = -\frac{4}{X^{n-1}}\qquad n > 1
}{eq:phase35}

Assuming in addition 
\eq{
U=-\frac{a}{X}
}{eq:phase36}
we can exploit the existing discussion of the so-called $ab$-family \cite{Katanaev:1997ni} reviewed in section 3.3 of \cite{Grumiller:2002nm} (see in particular their Fig.~3.12, which contains the ``phase space'' of all possible Penrose diagrams). The parameter $a$ is the same in both discussions and the parameters $b$ and $n$ are related by
$n = -b$.

With our assumptions of asymptotic mass-domination and existence of a well-defined variational principle we have to satisfy the inequalities
$a \le 1$, 
$b < -1$.
From Fig.~3.12 in \cite{Grumiller:2002nm} we then see that we always have the same Penrose diagram, namely that of the Schwarzschild black hole.

However, there is one catch: the constructions in \cite{Katanaev:1997ni,Grumiller:2002nm} assumed that the dilaton field depends on a spacelike (``radial'') coordinate. But in our case the dilaton field depends on a timelike coordinate, by assumption. As a consequence, the Penrose diagram has to be rotated by $\pi/2$.

In conclusion, for 2d (generalized) FSCs we end up with the Penrose diagram depicted in Figure~\ref{fig:penrose_diagram} above, where we used standard nomenclature ($\cal{H}^\pm$ denotes Killing horizons, $\cal{J}^\pm$ denotes lightlike asymptotic boundaries, $i^\pm$ denotes timelike asymptotic future and past, the wiggly lines are singularities in the causal structure; region I corresponds to an expanding FSC, region II to a contracting one; various ${\mathfrak t}=\rm const.$ slices are depicted by curved lines in regions I and II). Our Penrose diagram coincides with the one in Fig.~2 by Cornalba and Costa \cite{Cornalba:2002fi}. 

\end{appendix}


 \providecommand{\href}[2]{#2}\begingroup\raggedright\endgroup

\end{document}